\documentclass{article}
\pdfoutput=1

\usepackage{amsfonts}
\usepackage{amssymb}
\usepackage{bbm}
\usepackage{bm}
\usepackage{booktabs}
\usepackage{caption}
\usepackage{empheq}
\usepackage[shortlabels,inline]{enumitem}
\usepackage{fancyhdr}
\usepackage{footnote}
\usepackage[T1]{fontenc}
\usepackage[margin=1in]{geometry}
\usepackage{graphicx}
\usepackage{mathtools}
\usepackage[numbers,sort&compress]{natbib}
\usepackage{subcaption}
\usepackage{titling}
\usepackage{titlesec}
\usepackage{titletoc}
\usepackage{xcolor}

    \usepackage{siunitx}
    \usepackage[plainpages=false,pdfpagelabels]{hyperref}
    \hypersetup{
        colorlinks=true,
        urlcolor=blue,
        linkcolor=blue,
        citecolor=blue
    }

    \usepackage[nameinlink,capitalize]{cleveref}
    \crefname{figure}{Figure}{Figures}
    \labelcrefformat{subequation}{#2(#1)#3}
    \labelcrefrangeformat{subequation}{#3(#1)#4 to #5(#2)#6}

    \let\originalleft\left
    \let\originalright\right
    \renewcommand*{\left}{\mathopen{}\mathclose\bgroup\originalleft}
    \renewcommand*{\right}{\aftergroup\egroup\originalright}

    \DeclarePairedDelimiter\parens{\lparen}{\rparen}
    \DeclarePairedDelimiter\bracks{\lbrack}{\rbrack}
    \DeclarePairedDelimiter\braces{\lbrace}{\rbrace}
    \DeclarePairedDelimiter\abs{\lvert}{\rvert}
    \DeclarePairedDelimiter\norm{\lVert}{\rVert}

    \newcommand*{\D}[2][]{\ensuremath\frac{\mathop{}\!\mathrm{d} #1}{\mathop{}\!\mathrm{d} #2}}
    \newcommand*{\grad}{\ensuremath\nabla}

    \newcommand*{\diff}{\mathop{}\!\mathrm{d}}

    \newcommand*{\mat}[2][b]{\ensuremath\begin{#1matrix}#2\end{#1matrix}}
    \newcommand*{\V}[1]{\ensuremath\bm{\mathrm{#1}}}

    \newcommand*{\schwarzR}{\ensuremath r_\text{S}}

    \newcommand*{\realSchwarzR}{\ensuremath R_\text{S}}

    \newcommand*{\realHorizonR}{\ensuremath R_\text{h}}
    \newcommand*{\realHorizonRDD}{\ensuremath P_\text{h}}

    \newcommand*{\spacetimeR}{\ensuremath r}
    \newcommand*{\spacetimeRDD}{\ensuremath \rho}
    \newcommand*{\spacetimeTheta}{\ensuremath \theta}
    \newcommand*{\spacetimePhi}{\ensuremath \phi}
    \newcommand*{\realR}{\ensuremath R}
    \newcommand*{\realRDD}{\ensuremath P}
    \newcommand*{\realTheta}{\ensuremath \Theta}
    \newcommand*{\realPhi}{\ensuremath \Phi}

    \newcommand*{\realX}{\ensuremath X}
    \newcommand*{\realXDD}{\ensuremath \hat{X}}
    
    \newcommand*{\realYDD}{\ensuremath \hat{Y}}
    
    \newcommand*{\realZDD}{\ensuremath \hat{Z}}

    \newcommand*{\spacetimeIsoRDD}{\ensuremath \tilde{\rho}}
    
    \newcommand*{\spacetimeIsoXDD}{\ensuremath \hat{x}}
    
    \newcommand*{\spacetimeIsoYDD}{\ensuremath \hat{y}}
    
    \newcommand*{\spacetimeIsoZDD}{\ensuremath \hat{z}}

    \newcommand*{\spacetimeTime}{\ensuremath t}
    \newcommand*{\spacetimeTimeDD}{\ensuremath \hat{t}}
    \newcommand*{\affine}{\ensuremath \sigma}
    \newcommand*{\affineDD}{\ensuremath \hat{\sigma}}
    \newcommand*{\spacetimeArcLength}{\ensuremath s}
    \newcommand*{\spacetimeArcLengthDD}{\ensuremath \hat{s}}

    \newcommand*{\realTimeDD}{\ensuremath \hat{T}}

    \newcommand*{\spacetimeEnergy}{\ensuremath \varepsilon}
    \newcommand*{\spacetimeAngMom}{\ensuremath \ell}
    
    \newcommand*{\spacetimeEnergyDD}{\ensuremath \hat{\varepsilon}}
    \newcommand*{\spacetimeAngMomDD}{\ensuremath \hat{\ell}}

    \newcommand*{\realEnergy}{\ensuremath \cE}
    \newcommand*{\realAngMom}{\ensuremath L}

    \newcommand*{\electricField}{\ensuremath E}
    \newcommand*{\magneticField}{\ensuremath B}
    \newcommand*{\displacementField}{\ensuremath D}
    \newcommand*{\magnetizingField}{\ensuremath H}
    \newcommand*{\magnetoelectric}{\ensuremath \alpha}
    \newcommand*{\permittivityTensor}{\ensuremath \epsilon}
    \newcommand*{\permeabilityTensor}{\ensuremath \mu}
    \newcommand*{\refIndex}{\ensuremath n}

    \newcommand*{\metricTensor}{\ensuremath g}
    \newcommand*{\coordMetric}{\ensuremath \gamma}

    \newcommand*{\bhMass}{\ensuremath M}
    \newcommand*{\bhAngMom}{\ensuremath J}
    \newcommand*{\knA}{\ensuremath a}
    \newcommand*{\knADD}{\ensuremath \hat{a}}
    \newcommand*{\bhCharge}{\ensuremath Q}
    \newcommand*{\bhChargeDD}{\ensuremath \hat{Q}}
    \newcommand*{\bhMagCharge}{\ensuremath Q_\text{m}}
    \newcommand*{\bhMagChargeDD}{\ensuremath \hat{Q}_\text{m}}
    
    \newcommand*{\knChargeRadiusDD}{\ensuremath \rho_Q}
    
    \newcommand*{\knSigmaDD}{\ensuremath \hat{\Sigma}}
    
    \newcommand*{\knDeltaDD}{\ensuremath \hat{\Delta}}
    
    \newcommand*{\knGeoVDD}{\ensuremath \hat{V}}

    \newcommand*{\spacetimeImpactDD}{\ensuremath \hat{b}}
    \newcommand*{\spacetimeImpactInf}{\ensuremath b_\infty}
    \newcommand*{\spacetimeImpactInfDD}{\ensuremath \hat{b}_\infty}
    \newcommand*{\trajAngle}{\ensuremath \beta}

    \newcommand*{\realImpact}{\ensuremath B}
    \newcommand*{\realImpactDD}{\ensuremath \hat{B}}

    \newcommand*{\freq}{\ensuremath f}
    
    \newcommand*{\wavelength}{\ensuremath \lambda}

    \newcommand*{\GaussWidth}{\delta}

    \newcommand*{\bbone}{\mathbbm{1}}
    \newcommand*{\cE}{\mathcal{E}}
    \DeclareMathOperator{\re}{Re}
    \DeclareMathOperator{\sgn}{sgn}

    \newcommand*{\KN}{Kerr--Newman}
    \newcommand*{\RN}{Reissner--Nordstr\"om}
    \newcommand*{\BL}{Boyer--Lindquist}

    \newcommand*{\tempWidth}{\textwidth}

    \newcommand{\final}[1]{#1}

    \makeatletter
    \def\blfootnote{\gdef\@thefnmark{}\@footnotetext}
    \makeatother

    \allowdisplaybreaks

    \title{Optical analogues to the equatorial \KN{} black hole}
    \author{R. A. Tinguely$^\dag$ and Andrew P. Turner$^\ddag$ \\[0.9em]
    \small\textit{$^\dag$Plasma Science and Fusion Center} \\
    \small\textit{Massachusetts Institute of Technology} \\
    \small\textit{77 Massachusetts Avenue} \\
    \small\textit{Cambridge, MA 02139, USA} \\[1em]
    \small\textit{$^\ddag$Center for Theoretical Physics} \\ 
    \small\textit{Massachusetts Institute of Technology} \\
    \small\textit{77 Massachusetts Avenue} \\
    \small\textit{Cambridge, MA 02139, USA}}
    \date{\today}

\begin{document}
\maketitle
\thispagestyle{fancy}

\begin{abstract}
    \final{Optical analogues to black holes allow the investigation of general relativity in a laboratory setting. Previous works have considered analogues to Schwarzschild black holes in an isotropic coordinate system; the major drawback is that required material properties diverge at the horizon. We present the dielectric permittivity and permeability tensors that exactly reproduce the equatorial \KN{} metric, as well as the gradient-index material that reproduces equatorial \KN{} null geodesics. Importantly, the radial profile of the scalar refractive index is finite along all trajectories except at the point of rotation reversal for counter-rotating geodesics. Construction of these analogues is feasible with available ordinary materials. A finite-difference frequency-domain solver of Maxwell's equations is used to simulate light trajectories around a variety of \KN{} black holes. For reasonably sized experimental systems, ray tracing confirms that null geodesics can be well-approximated in the lab, even when allowing for imperfect construction and experimental error.}
\end{abstract}

\blfootnote{
$^\dag$\texttt{rating} at \texttt{mit.edu},
$^\ddag$\texttt{apturner} at \texttt{mit.edu}
}
\vspace{-1.2em}

\section*{}\label{sec:intro}
    In recent years, there has been a great amount of interest in precisely controlling the electromagnetic response of artificial materials. By introducing subwavelength structural features, the permittivity and permeability tensors of the medium can be tuned to exhibit a wide range of interesting and useful phenomena, such as cloaking \cite{pendry2006,leonhardt2006,leonhardt2006notes,schurig2006,cai2007,chen2007,li2008}, negative refraction \cite{pendry2006,smith2004,veselago2006}, and subwavelength microscopy with superlenses \cite{grbic2004,fang2005,lee2005,melville2005}.

    Analogue spacetimes \cite{leonhardt2006,pendry2006,leonhardt2006general,leonhardt2009,chen2010transformation,leonhardt2010,xu2015,plebanski1960} use optical materials to implement coordinate transformations between a physical space and a virtual ``electromagnetic space,'' via the formal equivalence between Maxwell's equations in curved spacetime and those in flat spacetime within a corresponding bianisotropic medium \cite{plebanski1960,felice1971,mashhoon1973,scully1982,thompson2010}. This allows one to build optical analogues to gravitational systems \cite{reznik2000,schutzhold2002,unruh2003,greenleaf2007,genov2009,narimanov2009,anderson2010,cheng2010,kildishev2010,li2010,mackay2010,smolyaninov2010,mackay2011,smolyaninov2011,wang2011,smolyaninov2012,yang2012,sheng2013,yin2013,bekenstein2015nature,patsyk2018}. In particular, there has been a fair amount of interest in reproducing the metrics of black holes \cite{leonhardt2003,chen2010,fernandeznunez2016,bekenstein2017,pires2018,zhou2019}. The null geodesics and polarizations of light moving in the spacetime metric can be reproduced exactly within a fully bianisotropic material; if one simply wishes to reproduce the null geodesics of the metric, however, it is much simpler to use an appropriately designed gradient-index material that is easier to construct experimentally.

    In this paper, we discuss the bianisotropic and gradient-index materials that imitate the exterior \final{equatorial} \KN{} black hole solution. \final{We} first carry out the analysis for optical systems reproducing the null geodesics of the Schwarzschild black hole. We recover the familiar results for the permittivity and permeability tensors and scalar refractive index reproducing the metric in isotropic coordinates, as well as the permittivity and permeability tensors reproducing the metric in the Schwarzschild coordinates \cite{felice1971,ye2008,chen2010,fernandeznunez2016}. We then present the scalar index that reproduces the null geodesics for Schwarzschild coordinates, which, by comparison with the isotropic result, has the significant experimental benefit of remaining finite all the way to the horizon. \final{We then} carry out these same analyses for the equatorial \KN{} metric in \BL{} coordinates, reproducing the metric within a fully bianisotropic material \cite{thompson2010}, and finding the scalar index required to reproduce the null geodesics. \final{We use} finite-difference frequency-domain simulations of systems that approximate the gradient-index solutions of the Schwarzschild and \KN{} black holes with concentric circular shells of constant index, and use ray tracing to perform an analysis of the error sensitivity of such systems. These analyses demonstrate that these approximate gradient-index systems, which are far simpler to construct than true gradient-index systems or full bianisotropic media, can adequately reproduce null geodesics and are forgiving to fabrication and experimental error for reasonable geodesics. As such, they are practical tabletop analogues for charged and/or rotating black holes.

\section*{Results}\label{sec:results}
    \final{
    Throughout this paper we use Gaussian Planck units, with $c = \hbar = G = 4 \pi \epsilon_0 = 1$. Greek indices range over temporal and spatial coordinates, e.g., $\mu = 0, \dots, 3$, while Roman indices range over only spatial coordinates, e.g., $i = 1, \dots, 3$. We use uppercase Greek and Roman letters to indicate variables related to the optical system, while we use lowercase letters to indicate variables related to the spacetime metric it is replicating. We refer to these respectively as ``real space'' and ``spacetime'' variables. We typically use hats to indicate the dimensionless versions of variables. When we map spacetime coordinates onto real space coordinates, we always do so by equating the dimensionless coordinates. Spacetime variables are dedimensionalized via multiplication by the appropriate power of the black hole mass $\bhMass$. Real space dimensionless variables are then dimensionalized by a convenient length scale for construction. Using this matching of coordinates allows one to more easily keep track of the relationship between real space coordinates and the spacetime coordinates they represent.
    }

    \subsection*{The Schwarzschild black hole}\label{sec:schwarz}
        We will begin by studying the Schwarzschild black hole and various optical analogues thereof. The Schwarzschild metric describes the spacetime geometry of a static, uncharged black hole of mass $\bhMass$, and is given in dimensionless Schwarzschild coordinates $\spacetimeArcLengthDD, \spacetimeTimeDD, \spacetimeRDD, \spacetimeTheta, \spacetimePhi$ (related to the usual dimensionful quantities via $\spacetimeArcLength = \bhMass \spacetimeArcLengthDD$, $\spacetimeTime = \bhMass \spacetimeTimeDD$, $\spacetimeR = \bhMass \spacetimeRDD$) by \cite{weinbergBook}
            \begin{equation}
                \label{eq:ddSchwarzMetric}
                \diff\spacetimeArcLengthDD^2 = -\parens*{1 - \frac{2}{\spacetimeRDD}} \diff\spacetimeTimeDD^2 + \parens*{1 - \frac{2}{\spacetimeRDD}}^{-1} \diff\spacetimeRDD^2 + \spacetimeRDD^2 \parens*{\diff\spacetimeTheta^2 + \sin^2\spacetimeTheta \diff\spacetimePhi^2}\,.
            \end{equation}

        Making the coordinate transformation $\spacetimeRDD = \spacetimeIsoRDD \parens*{1 + \frac{1}{2 \spacetimeIsoRDD}}^2$, the Schwarzschild metric \labelcref{eq:ddSchwarzMetric} can be written in the form \cite{weinbergBook}
            \begin{equation}
                \label{eq:ddSchwarzIsoMetric}
                \diff\spacetimeArcLengthDD^2 = -\frac{\parens*{1 - \frac{1}{2 \spacetimeIsoRDD}}^2}{\parens*{1 + \frac{1}{2 \spacetimeIsoRDD}}^2} \diff\spacetimeTimeDD^2 + \parens*{1 + \frac{1}{2 \spacetimeIsoRDD}}^4 \parens*{\diff\spacetimeIsoXDD^2 + \diff\spacetimeIsoYDD^2 + \diff\spacetimeIsoZDD^2}\,,
            \end{equation}
        where the spacetime isotropic coordinates $\parens*{\spacetimeIsoXDD, \spacetimeIsoYDD, \spacetimeIsoZDD}$ are related to the transformed Schwarzschild coordinates $\parens*{\spacetimeIsoRDD, \spacetimeTheta, \spacetimePhi}$ via the transformation from Cartesian to spherical coordinates.

        We \final{first replicate} the metric in isotropic coordinates, given in \cref{eq:ddSchwarzIsoMetric}, in order to make contact with existing literature. As discussed in \cite{plebanski1960,leonhardt2009}, there is a formal equivalence between the equations of electrodynamics in a curved spacetime and those in flat space in a macroscopic medium. Specifically, the behavior of light in a curved spacetime background described by metric $\metricTensor_{\mu \nu}$ is reproduced in flat space within an impedance-matched bianisotropic medium with permittivity $\permittivityTensor^{i j}$, permeability $\permeabilityTensor^{i j}$, and magnetoelectric coupling $\magnetoelectric_i$ given by
            \begin{equation}
                \label{eq:analogueEpsMuAl}
                \permittivityTensor^{i j} = \permeabilityTensor^{i j} = -\frac{\sqrt{-\det\metricTensor}}{\metricTensor_{0 0} \sqrt{\det\coordMetric}} \metricTensor^{i j}\,, \quad \magnetoelectric_i = \frac{\metricTensor_{0 i}}{\metricTensor_{0 0} \sqrt{\det\coordMetric}}\,,
            \end{equation}
        where $\coordMetric_{i j}$ is the three-dimensional metric tensor of the real space coordinate system in which we construct the medium, onto which we map the spatial components $\metricTensor_{i j}$. Here, $\metricTensor$ and $\coordMetric$ denote the determinants of $\metricTensor_{\mu \nu}$ and $\coordMetric_{i j}$, respectively. The macroscopic fields $\V{\displacementField}, \V{\magnetizingField}$ are related to the microscopic fields $\V{\electricField}, \V{\magneticField}$ via
            \begin{equation}
                \V{\displacementField} = \bm{\permittivityTensor} \V{\electricField} + \V{\magnetoelectric} \times \V{\magnetizingField}\,, \quad \V{\magneticField} = \bm{\permeabilityTensor} \V{\magnetizingField} - \V{\magnetoelectric} \times \V{\electricField}\,.
            \end{equation}
        As discussed in \cite{fathi2016}, this choice of identification between the spacetime geometry and the electromagnetic analogue, elaborated first in \cite{plebanski1960}, is not unique, and cannot reproduce all measurable properties of light moving in the spacetime metric. However, it is sufficient to reproduce both the null geodesic trajectory and the polarizations of light moving along these geodesics, which makes analogues produced with this identification worthy subjects of study.

        Using \cref{eq:analogueEpsMuAl} to map the dimensionless spacetime isotropic coordinates $\parens*{\spacetimeIsoXDD, \spacetimeIsoYDD, \spacetimeIsoZDD}$ onto the corresponding dimensionless real space Cartesian coordinates $\parens*{\realXDD, \realYDD, \realZDD}$ (and thus mapping the dimensionless spacetime isotropic radial coordinate $\spacetimeIsoRDD$ onto the dimensionless real space radial coordinate $\realRDD$), we find that the behavior of light in the Schwarzschild metric \labelcref{eq:ddSchwarzIsoMetric} is reproduced in flat space within a medium described by
            \begin{equation}
                \label{eq:schwarzIsoTens}
                \permittivityTensor^{i j} = \permeabilityTensor^{i j} = \frac{(2 \realRDD + 1)^3}{4 \realRDD^2 (2 \realRDD - 1)} \bbone^{i j}\,, \quad i, j \in \braces*{\realXDD, \realYDD, \realZDD}\,.
            \end{equation}
        In this case, the medium is isotropic, and the scalar index can be read off immediately from \cref{eq:schwarzIsoTens} as
            \begin{equation}
                \label{eq:schwarzIsoIndex}
                \refIndex(\realRDD) = \frac{(2 \realRDD + 1)^3}{4 \realRDD^2 (2 \realRDD - 1)}\,.
            \end{equation}
        Note that the results of \cref{eq:schwarzIsoTens,eq:schwarzIsoIndex} are well-established in the literature \cite{felice1971,ye2008,chen2010,fernandeznunez2016}. \Cref{eq:schwarzIsoIndex} has the benefit that there is a single scalar index that reproduces all null geodesics and the polarization of light moving along these geodesics, and \final{as} the material is isotropic (though still inhomogeneous) \final{it is} thus easier to construct experimentally. However, this refractive index diverges approaching the horizon, i.e., as $\realRDD \to \frac{1}{2}$, so it is not useful for investigating geodesics in the vicinity of the horizon.

        Another approach is to instead use the Schwarzschild coordinates \labelcref{eq:ddSchwarzMetric}, which produces an anisotropic medium distinct from \cref{eq:schwarzIsoTens}. As before, we use \cref{eq:analogueEpsMuAl} to map the dimensionless spacetime Schwarzschild coordinates $\parens*{\spacetimeRDD, \spacetimeTheta, \spacetimePhi}$ (now with the Schwarzschild radial coordinate $\spacetimeRDD$ rather than the isotropic radial coordinate $\spacetimeIsoRDD$) onto the corresponding dimensionless real space spherical coordinates $\parens*{\realRDD, \realTheta, \realPhi}$, yielding
            \begin{equation}
                \label{eq:schwarzSpherTens}
                \permittivityTensor^{i j} = \permeabilityTensor^{i j} = \mat[p]{1 & 0 & 0 \\ 0 & \frac{1}{\realRDD (\realRDD - 2)} & 0 \\ 0 & 0 & \frac{\csc(\realTheta)^2}{\realRDD (\realRDD - 2)}}\,, \quad i, j \in \braces*{\realRDD, \realTheta, \realPhi}\,.
            \end{equation}
        This same system is described in dimensionless real space Cartesian coordinates $\parens*{\realXDD, \realYDD, \realZDD}$ by
            \begin{equation}
                \label{eq:schwarzCartTens}
                \permittivityTensor^{i j} = \permeabilityTensor^{i j} = \frac{1}{\realRDD^2 (2 - \realRDD)} \mat[p]{2 \realXDD^2 - \realRDD^3 & 2 \realXDD \realYDD & 2 \realXDD \realZDD \\ 2 \realXDD \realYDD & 2 \realYDD^2 - \realRDD^3 & 2 \realYDD \realZDD \\ 2 \realXDD \realZDD & 2 \realYDD \realZDD & 2 \realZDD^2 - \realRDD^3} = \frac{2 \realRDD^i \realRDD^j - \realRDD^3 \bbone^{i j}}{\realRDD^2 (2 - \realRDD)}\,, \quad i, j \in \braces*{\realXDD, \realYDD, \realZDD}\,,
            \end{equation}
        where $\realRDD^i = \parens*{\realXDD, \realYDD, \realZDD}$ and the dimensionless real space Cartesian coordinates $\parens*{\realXDD, \realYDD, \realZDD}$ are related to the real space spherical coordinates $\parens*{\realRDD, \realTheta, \realPhi}$ in the usual way. This result matches those presented in \cite{chen2010,fernandeznunez2016}.

        If we only wish to reproduce the trajectories of light in the Schwarzschild metric (and not the proper polarizations), then a radially varying scalar index $\refIndex(\realRDD)$ is sufficient. In Schwarzschild coordinates, we will find that the radial profile depends on the initial conditions \final{defining the geodesic}. We consider null geodesics of the metric \labelcref{eq:ddSchwarzMetric}; all such geodesics are planar, and so the spherical symmetry allows us to take $\theta = \pi / 2$ without loss of generality. Such null geodesics of the Schwarzschild metric are parametrized by a conserved energy at infinity, $\spacetimeEnergy = \parens*{1 - \frac{2 \bhMass}{\spacetimeR}} \D[\spacetimeTime]{\affine}$, and the conserved angular momentum, $\spacetimeAngMom = \spacetimeR^2 \D[\spacetimePhi]{\affine}$, with $\affine$ the affine parameter of the geodesic. Dedimensionalizing these parameters via $\spacetimeAngMom = \bhMass \spacetimeAngMomDD$ and $\affine = \bhMass \affineDD$ (note that the energy is already dimensionless, $\spacetimeEnergyDD = \spacetimeEnergy$), null geodesics satisfy the geodesic equation \cite{weinbergBook}
            \begin{equation}
                \label{eq:schwarzSpherGeod}
                -\spacetimeEnergyDD^2 + \parens*{\D[\spacetimeRDD]{\affineDD}}^2 + \parens*{1 - \frac{2}{\spacetimeRDD}} \frac{\spacetimeAngMomDD^2}{\spacetimeRDD^2} = 0\,.
            \end{equation}
        Combining this equation with $\parens*{\D[\spacetimePhi]{\affineDD}}^2 = \spacetimeAngMomDD^2 / \spacetimeRDD^4$ yields
            \begin{equation}
                \label{eq:schwarzDphidr}
                \D[\spacetimePhi]{\spacetimeRDD} = \pm\parens*{\frac{\spacetimeEnergyDD^2}{\spacetimeAngMomDD^2} \spacetimeRDD^4 - \spacetimeRDD^2 + 2 \spacetimeRDD}^{-1 / 2}\,.
            \end{equation}
        We then make use of the spacetime impact parameter $\spacetimeImpactDD(\spacetimeRDD) = \spacetimeRDD \sin\trajAngle$, where $\trajAngle$ is defined by the relation
            \begin{equation}
                \label{eq:spacetimeTrajAngle}
                \spacetimeRDD \D[\spacetimePhi]{\spacetimeRDD} = -\tan\trajAngle\,.
            \end{equation}
        Plugging this relation into \cref{eq:schwarzDphidr} and making the sign choice consistent with our definition of $\trajAngle$, we find that
            \begin{equation}
                \label{eq:schwarzImpactDD}
                \spacetimeImpactDD(\spacetimeRDD) = \parens*{\spacetimeImpactInfDD^{-2} + 2 \spacetimeRDD^{-3}}^{-1 / 2}\,,
            \end{equation}
        where we have defined $\spacetimeImpactInfDD = \spacetimeAngMomDD / \spacetimeEnergyDD$. Fermat's principle relates the real space impact parameter and index of refraction by
            \begin{equation}
                \label{eq:fermatIndexDD}
                \refIndex(\realRDD) \propto \realImpactDD(\realRDD)^{-1}\,.
            \end{equation}
        \Cref{eq:schwarzImpactDD} is then taken as input to \cref{eq:fermatIndexDD} by equating the spacetime coordinates $\parens*{\spacetimeRDD, \spacetimePhi}$ with the real space coordinates $\parens*{\realRDD, \realPhi}$, which also equates the dimensionless spacetime impact parameter $\spacetimeImpactDD$ with the dimensionless real space impact parameter $\realImpactDD$. This yields
            \begin{equation}
                \label{eq:schwarzIndex}
                \refIndex(\realRDD) \propto \sqrt{\spacetimeImpactInfDD^{-2} + 2 \realRDD^{-3}}\,.
            \end{equation}
        This solution has a number of noteworthy features. First, we reiterate that \cref{eq:schwarzIndex} only reproduces the geodesic trajectories of light moving in the Schwarzschild metric \labelcref{eq:ddSchwarzMetric}, but does not faithfully reproduce its polarizations. The radial profile depends on the initial condition $\spacetimeImpactInfDD$, which is related to initial angle $\trajAngle$ and initial radius $\realRDD_0$ by
            \begin{equation}
                \label{eq:impactToAngle}
                \spacetimeImpactInfDD^2 = \frac{\realRDD_0^2}{\csc^2\trajAngle_0 - 2 \realRDD_0^{-1}}\,.
            \end{equation}
        This is somewhat inconvenient for experimental application, as it means that a different apparatus must be constructed for each family of geodesics; to address this, one could in principle construct a cylinder, where $\spacetimeImpactInfDD$ varies along the cylinder axis and each 2D slice recreates the corresponding family of null geodesics. A significant benefit of this coordinate system, however, is that $\refIndex(\realRDD)$ approaches a finite value as $\realRDD \to 2$ so long as $\spacetimeImpactInfDD \ne 0$, which means that geodesics can be studied in the vicinity of the event horizon, in contrast to the solution \labelcref{eq:schwarzIsoIndex}. The constant of proportionality in \cref{eq:schwarzIndex} allows us to tune the scalar index at the initial $\realRDD_0$ to the most feasible value for construction.

        Finally, note that we can relate conserved quantities $\spacetimeEnergy$, $\spacetimeAngMom$ in spacetime to $\realEnergy, \realAngMom$ in real space in the following way: in flat space, a photon with frequency $\freq$ and wavelength $\wavelength$ has energy $\realEnergy = 2 \pi \freq$ and angular momentum $\realAngMom = 2 \pi \realImpact / \wavelength$, with $\realImpact$ the dimensionful real space impact parameter. Thus, $\realAngMom / \realEnergy \realSchwarzR = \refIndex \realImpactDD / 2 = \text{constant}$. Setting $\refIndex = 1$ at $\realRDD \to \infty$ in \cref{eq:schwarzIndex} and equating the real space and spacetime dimensionless impact parameters, $\realImpactDD = \spacetimeImpactDD$, yields $\realAngMom / \realEnergy \realSchwarzR = \spacetimeAngMom / \spacetimeEnergy \schwarzR$. Here, $\schwarzR = 2 \bhMass$, and $\realSchwarzR$ is the real space radius onto which $\schwarzR$ is mapped.

    \subsection*{The \KN{} black hole}\label{sec:kerr}
        We now apply the same approaches to investigate optical analogues of the \KN{} black hole, of which the Kerr, \RN{}, and Schwarzschild results are special cases. We will restrict our attention to equatorial null geodesics.

        The \KN{} metric describes the spacetime geometry surrounding a black hole of mass $\bhMass$, angular momentum per unit mass $\knA = \bhAngMom / \bhMass$, electric charge $\bhCharge$, and magnetic charge $\bhMagCharge$. Dedimensionalizing the quantities via $\knA = \bhMass \knADD$, $\bhCharge = \bhMass \bhChargeDD$, $\bhMagCharge = \bhMass \bhMagChargeDD$, the metric is given in dimensionless \BL{} coordinates by \cite{weinbergBook}
            \begin{equation}
                \label{eq:ddBLMetric}
                \diff\spacetimeArcLengthDD^2 = \knSigmaDD \parens*{\frac{\diff\spacetimeRDD^2}{\knDeltaDD} + \diff\spacetimeTheta^2} - \frac{\knDeltaDD}{\knSigmaDD} \parens*{\diff\spacetimeTimeDD - \knADD \sin^2\spacetimeTheta \diff\spacetimePhi}^2  + \frac{\sin^2\spacetimeTheta}{\knSigmaDD} \bracks*{\parens*{\spacetimeRDD^2 + \knADD^2} \diff\spacetimePhi - \knADD \diff\spacetimeTimeDD}^2\,,
            \end{equation}
        where
            \begin{equation}
                \begin{aligned}
                    \knSigmaDD &= \spacetimeRDD^2 + \knADD^2 \cos^2\spacetimeTheta\,, \\
                    \knDeltaDD &= \spacetimeRDD^2 - 2 \spacetimeRDD + \knADD^2 + \knChargeRadiusDD^2\,, \\
                    \knChargeRadiusDD^2 &= \bhChargeDD^2 + \bhMagChargeDD^2\,.
                \end{aligned}
            \end{equation}
        Here, $\bhMass$ is the total mass-equivalent, which contains contributions from the irreducible mass, the rotational energy, and the Coulomb energy of the black hole \cite{christodoulou1971}.

        After setting $\spacetimeTheta = \pi / 2$ and $\diff\spacetimeTheta = 0$ to restrict to the equatorial case, we use \cref{eq:analogueEpsMuAl} to map the dimensionless spacetime coordinates $(\spacetimeRDD, \spacetimeTheta, \spacetimePhi)$ onto the dimensionless real space coordinates $(\realRDD, \realTheta, \realPhi)$, as before, yielding
            \begin{equation}
                \label{eq:kerrTens}
                \permittivityTensor^{i j} = \permeabilityTensor^{i j} = \mat[p]{\frac{\knDeltaDD}{\knDeltaDD - \knADD^2} & 0 & 0 \\ 0 & \frac{1}{\knDeltaDD - \knADD^2} & 0 \\ 0 & 0 & \frac{1}{\knDeltaDD}}\,, \quad \magnetoelectric_i = \mat[p]{0 \\ 0 \\ \knADD \parens*{\frac{1}{\knDeltaDD - \knADD^2} - \frac{1}{\realRDD^2}}}\,,
            \end{equation}
        where $\knDeltaDD$ should now be interpreted as a function of $\realRDD$. \final{The derivation is given in full in the Methods section.} The equatorial geodesics \final{and polarizations} of the \KN{} metric are exactly reproduced in flat space within a medium with the`'se properties \cite{thompson2010}. There is a subtlety here---although the radial and azimuthal components $\realRDD, \realTheta$ appear to diverge at the ergosphere $\knDeltaDD = \knADD^2$, this is a spurious divergence. As discussed in \cite{thompson2010,thompson2010-2,thompson2011}, the physically relevant covariant quantity is the tensor $\bm{\chi}$ \final{defined therein}, which relates the macroscopic and microscopic fields. This quantity diverges only at the horizon $\knDeltaDD = 0$.

        As before, we can also replicate equatorial null geodesics of the \KN{} metric using only a scalar index. As in the Schwarzschild case, these geodesics are parametrized by the dimensionless conserved energy at infinity and conserved angular momentum, given in this case by
            \begin{equation}
                \begin{aligned}
                    \spacetimeEnergyDD &= \parens*{1 - \frac{2}{\spacetimeRDD} + \frac{\knChargeRadiusDD^2}{\spacetimeRDD^2}} \D[\spacetimeTimeDD]{\affineDD} + \parens*{\frac{2 \knADD}{\spacetimeRDD} - \frac{\knChargeRadiusDD^2 \knADD}{\spacetimeRDD^2}} \D[\spacetimePhi]{\affineDD}\,, \\
                    \spacetimeAngMomDD &= -\parens*{\frac{2 \knADD}{\spacetimeRDD} - \frac{\knChargeRadiusDD^2 \knADD}{\spacetimeRDD^2}} \D[\spacetimeTimeDD]{\affineDD} + \parens*{\spacetimeRDD^2 + \knADD^2 + \frac{2 \knADD^2}{\spacetimeRDD} - \frac{\knChargeRadiusDD^2 \knADD^2}{\spacetimeRDD^2}} \D[\spacetimePhi]{\affineDD}\,.
                \end{aligned}
            \end{equation}
        The geodesic equations describing the equatorial motion are
            \begin{equation}
                \label{eq:geodesicKN}
                \begin{aligned}
                    \D[\spacetimePhi]{\affineDD} &= \frac{1}{\knDeltaDD} \bracks*{\parens*{1 - \frac{2}{\spacetimeRDD} + \frac{\knChargeRadiusDD^2}{\spacetimeRDD^2}} \spacetimeAngMomDD + \parens*{\frac{2 \knADD}{\spacetimeRDD} - \frac{\knChargeRadiusDD^2 \knADD}{\spacetimeRDD^2}} \spacetimeEnergy}\,, \\
                    \parens*{\D[\spacetimeRDD]{\affineDD}}^2 &= \frac{\bracks*{\parens*{\spacetimeRDD^2 + \knADD^2}^2 - \knADD^2 \knDeltaDD} \spacetimeAngMomDD^2}{\spacetimeRDD^4} \parens*{\spacetimeImpactInfDD^{-1} - \knGeoVDD_+} \parens*{\spacetimeImpactInfDD^{-1} - \knGeoVDD_-}\,,
                \end{aligned}
            \end{equation}
        where
            \begin{equation}
                \knGeoVDD_\pm = \frac{\knADD \parens*{2 \spacetimeRDD - \knChargeRadiusDD^2} \pm \sgn(\spacetimeAngMomDD) \spacetimeRDD^2 \sqrt{\knDeltaDD}}{\parens*{\spacetimeRDD^2 + \knADD^2}^2 - \knADD^2 \knDeltaDD}\,.
            \end{equation}

        As before, we find the impact parameter $\spacetimeImpactDD(\spacetimeRDD) = \spacetimeRDD \sin\trajAngle$ by plugging \cref{eq:spacetimeTrajAngle} into $\D[\spacetimePhi]{\spacetimeRDD} = \frac{\mathrm{d}\spacetimePhi / \mathrm{d}\affineDD}{\mathrm{d}\spacetimeRDD / \mathrm{d}\affineDD}$, which yields
            \begin{equation}
                \spacetimeImpactDD(\spacetimeRDD) = \frac{\spacetimeRDD^2 \bracks*{\parens*{\knDeltaDD - \knADD^2} + \parens*{2 \spacetimeRDD - \knChargeRadiusDD^2} \knADD \spacetimeImpactInfDD^{-1}}}{\sqrt{\spacetimeRDD^2 \bracks*{\parens*{\knDeltaDD - \knADD^2} + \parens*{2 \spacetimeRDD - \knChargeRadiusDD^2} \knADD \spacetimeImpactInfDD^{-1}}^2 + \knDeltaDD^2 \bracks*{\parens*{\spacetimeRDD^2 + \knADD^2}^2 - \knADD^2 \knDeltaDD} \parens*{\spacetimeImpactInfDD^{-1} - \knGeoVDD_+} \parens*{\spacetimeImpactInfDD^{-1} - \knGeoVDD_-}}}\,.
            \end{equation}
        We proceed by equating spacetime coordinates $(\spacetimeRDD, \spacetimeTheta, \spacetimePhi)$ and real space coordinates $(\realRDD, \realTheta, \realPhi)$, which sets $\realImpactDD(\realRDD) = \spacetimeImpactDD(\spacetimeRDD = \realRDD)$. The scalar index for an optical \KN{} black hole is again given by
            \begin{equation}
                \label{eq:knScalarRefIndex}
                n(\realRDD) \propto \realImpactDD(\realRDD)^{-1}\,.
            \end{equation}
        An optical system with this scalar index reproduces the equatorial null geodesic trajectories of the \KN{} metric.

        Unlike the Schwarzschild case, this scalar index is not always sufficient to fully reproduce the given family of \KN{} geodesics. This can be seen immediately by noting that initially counter-rotating geodesics (those with $\spacetimeAngMomDD$ of opposite sign to $\knADD$) must turn around and become co-rotating before crossing into the ergosphere; such a reversal of the sign of $\D[\spacetimePhi]{\spacetimeRDD}$ is not possible with a finite (and positive) scalar index. This shortcoming manifests itself as a divergence of the scalar index; the outermost divergence occurs at radius
            \begin{equation}
                \realRDD_* = 1 - \knADD\spacetimeImpactInfDD^{-1} + \sqrt{\parens*{1 - \knADD\spacetimeImpactInfDD^{-1}} \parens*{1 - \knADD\spacetimeImpactInfDD^{-1} - \knChargeRadiusDD^2}}\,.
            \end{equation}
        This is a removable pole in the Schwarzschild and \RN{} cases. For rotating black holes, the divergence occurs at the point in the trajectory where the direction of rotation reverses, consistent with the above observation that a finite radially varying scalar index is insufficient to implement such a reversal. Thus, the pole only affects initially counter-rotating geodesics that enter the ergosphere.

    \subsection*{Simulations of constructible optical black holes}\label{sec:sims}
        Optical analogues to black holes are particularly useful if their constructions are realizable. In the following sections, we model optical black holes with radially varying scalar refractive indices $\refIndex(\realRDD)$, as given by \cref{eq:schwarzIndex,eq:knScalarRefIndex}. For Schwarzschild (and many \KN{}) black holes, $\refIndex(\realRDD)$ is maximal at the horizon. Because the impact parameter of light on the optical black hole must be less than or equal to the radius of the ``edge'' of the system, i.e., $\realImpactDD \le \realRDD_0$, it is found that $\refIndex(\realRDD) \le c_0 \refIndex_0 \realRDD_0$, where $c_0 = \sqrt{31/108} \approx 0.54$ and $\refIndex_0 = \refIndex(\realRDD_0)$. Thus, the construction of an optical Schwarzschild black hole with $\refIndex_0 = 1$ and moderate $\realRDD_0 \le 6$ is plausible and achievable with indices of refraction in the range of ordinary materials such as water, glass, and plastic. (As will be seen, many optical \KN{} black holes are also constructible.) True gradient-index profiles of the form \labelcref{eq:schwarzIndex} could perhaps be achieved with metamaterials; however, it is not clear how easily realizable such systems are, so in this work we approximate the profiles with concentric annuli of constant scalar index.

        For a system size in which the wavelength of the source light is much smaller than the gradient length scale of the scalar-index profile, i.e., $\wavelength \ll \refIndex / \norm{\grad\refIndex}$, a highly localized and highly directional light source, like a laser, would nearly approximate the geodesics of \cref{eq:schwarzDphidr,eq:geodesicKN}. Simulating these trajectories amounts to ray tracing, which we pursue in the following section. Specifically, we investigate the number of annuli needed to sufficiently mimic the true scalar-index profile and explore the impact of imperfect construction and experimental error on the deviation of the ray trajectory from the geodesic. However, in the following section, we will first consider the case in which the source wavelength is similar to the size of the optical black hole, i.e., \final{$\bhMass/\wavelength \sim O(10)$.} This is done to demonstrate the strengths and limitations of this study's approach, as well as to be consistent with previous studies such as \cite{genov2009,narimanov2009,chen2010,cheng2010,lu2010,kildishev2010,wang2011,sheng2013,fernandeznunez2016}.

        In this study, all optical black holes are modeled with dimensionless outer radius $\realRDD_0 = 6$. The system comprises either 16 or 21 concentric annuli, with the number depending on acceptable annulus thickness (i.e., greater than the wavelength) and the minimum modeled radius $\realRDD_{\mathrm{min}}$. The innermost and outermost annuli each have half the width of each interior annulus. The scalar index of each annulus is uniform, so that the simulated $\refIndex(\realRDD)$ profiles are piecewise functions, as shown in \final{\cref{fig:nSchwarzschild,fig:nKerrNewman}}. The values of $\refIndex$ for the innermost and outermost annuli are taken as $\refIndex(\realRDD)$ at the minimum and maximum radii, respectively; the refractive index of an inner annulus is taken as the value of $\refIndex(\realRDD)$ at its center.

        It is important to note here that this geometry was chosen for simplicity in the finite-difference frequency-domain simulations of the next section, in which dimensions are constrained by the wavelength. For consistency, the same geometry is scaled linearly for the ray tracing analyses \final{that follow}. In practice, non-uniform annulus thicknesses could be used to minimize steps $\Delta\refIndex$ in regions of high $\D[\refIndex]{\realR}$ and to reduce light scattering at each boundary, but this is left for future work.

        \renewcommand{\tempWidth}{0.5\textwidth}
        \begin{figure}[ht!]
            \centering
            \includegraphics[width=\tempWidth]{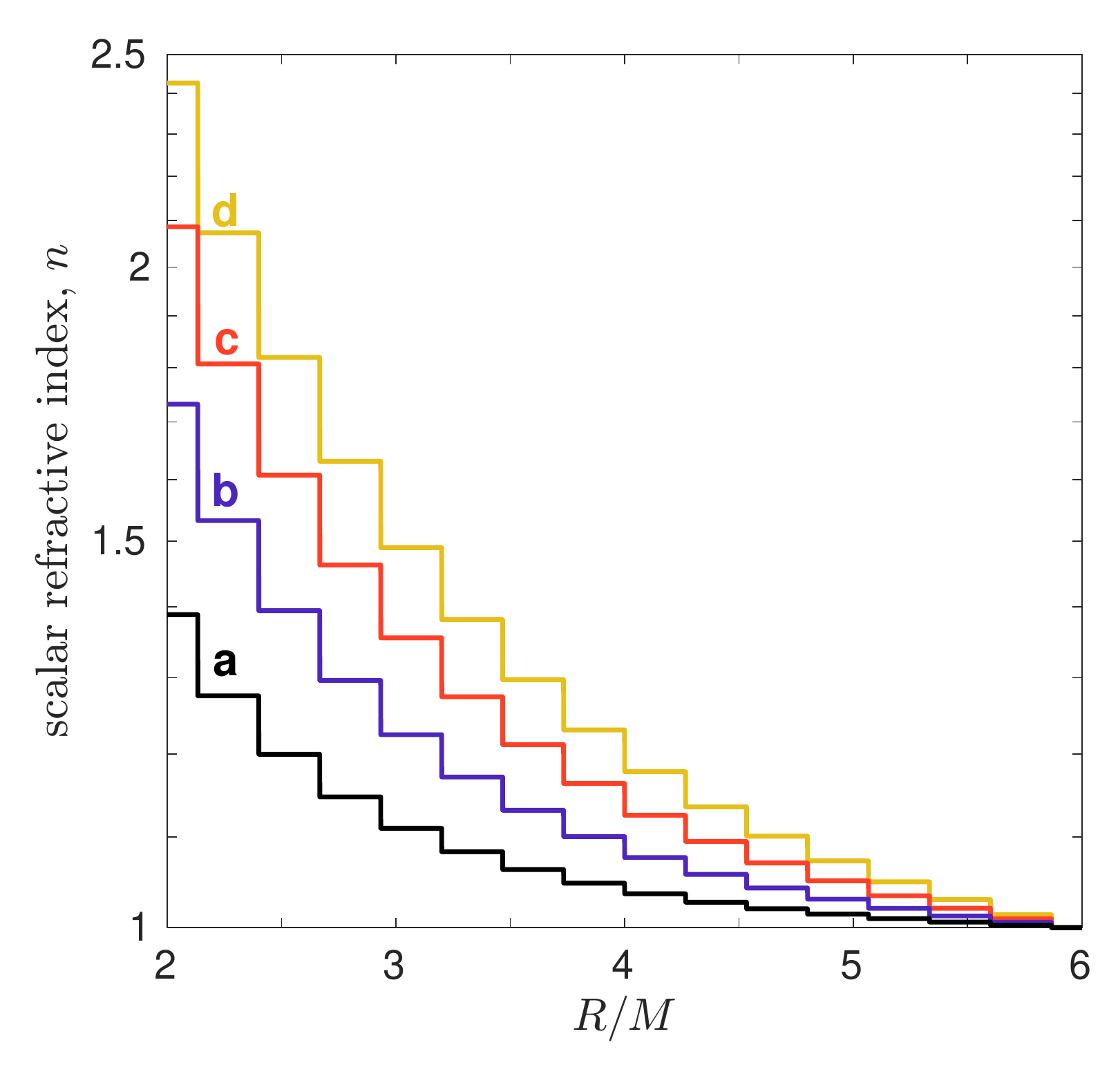}
            \caption{\final{\textbf{Scalar refractive index of simulated optical Schwarzschild black holes.} Radial profiles of the scalar refractive index used for simulations of optical Schwarzschild black holes with impact parameters $\spacetimeImpactInfDD =$~\textbf{a}~2, \textbf{b}~3, \textbf{c}~4, and \textbf{d}~5. The outer radius is $\realR_0/\bhMass = 6$ with $\bhMass$ the black hole mass. Note the logarithmic scale of the vertical axis.}}
            \label{fig:nSchwarzschild}
        \end{figure}

        \renewcommand{\tempWidth}{0.5\textwidth}
        \begin{figure}[ht!]
            \centering
            \includegraphics[width=\tempWidth]{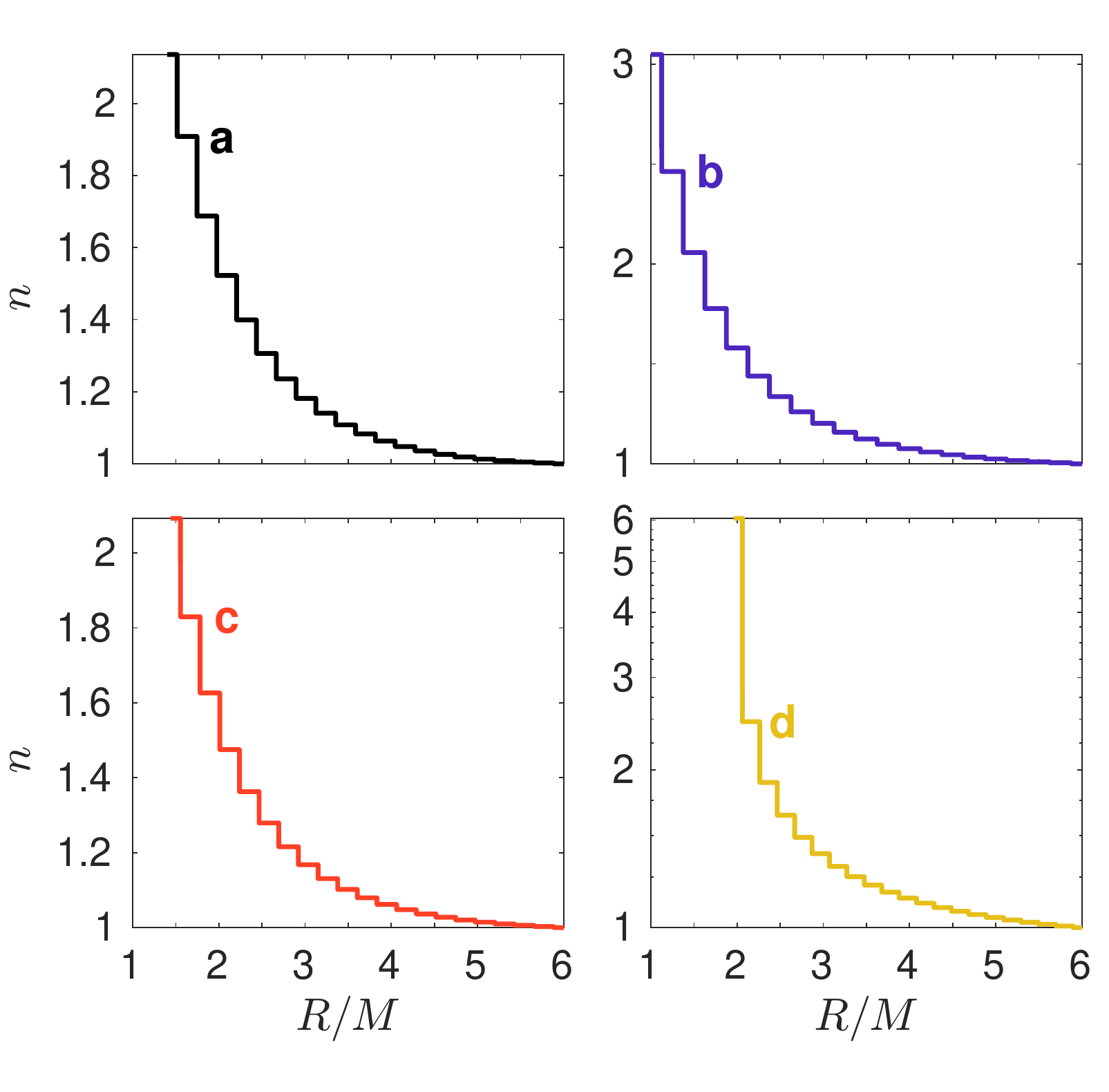}
            \caption{\final{\textbf{Scalar refractive index of simulated optical \KN{} black holes.} Radial profiles of the scalar refractive index $\refIndex$ used for simulations of optical \KN{} black holes: \textbf{a}~maximally co-rotating ($\knADD = 1, \knChargeRadiusDD = 0$); \textbf{b}~maximally charged ($\knADD = 0, \knChargeRadiusDD = 1$); \textbf{c}~charged and co-rotating ($\knADD = 2 / 5, \knChargeRadiusDD = 4 / 5$); and \textbf{d}~charged and counter-rotating ($\knADD = -2 / 5, \knChargeRadiusDD = 4 / 5$). The impact parameter is $\spacetimeImpactInfDD = 3$, and outer radius is $\realR_0/\bhMass = 6$, with $\bhMass$ the black hole mass. Note the logarithmic scales and limits of the vertical axes.}}
            \label{fig:nKerrNewman}
        \end{figure}

    \subsection*{Finite-difference frequency-domain simulations}\label{sec:fdfd}

\begin{figure}[h!]
            \centering
            \begin{subfigure}{\textwidth}
                \phantomcaption{}
                \label{fig:bInf10}
                \phantomcaption{}
                \label{fig:bInf15}
                \phantomcaption{}
                \label{fig:bInf20}
                \phantomcaption{}
                \label{fig:bInf25}
            \end{subfigure}
            \includegraphics[width=\textwidth]{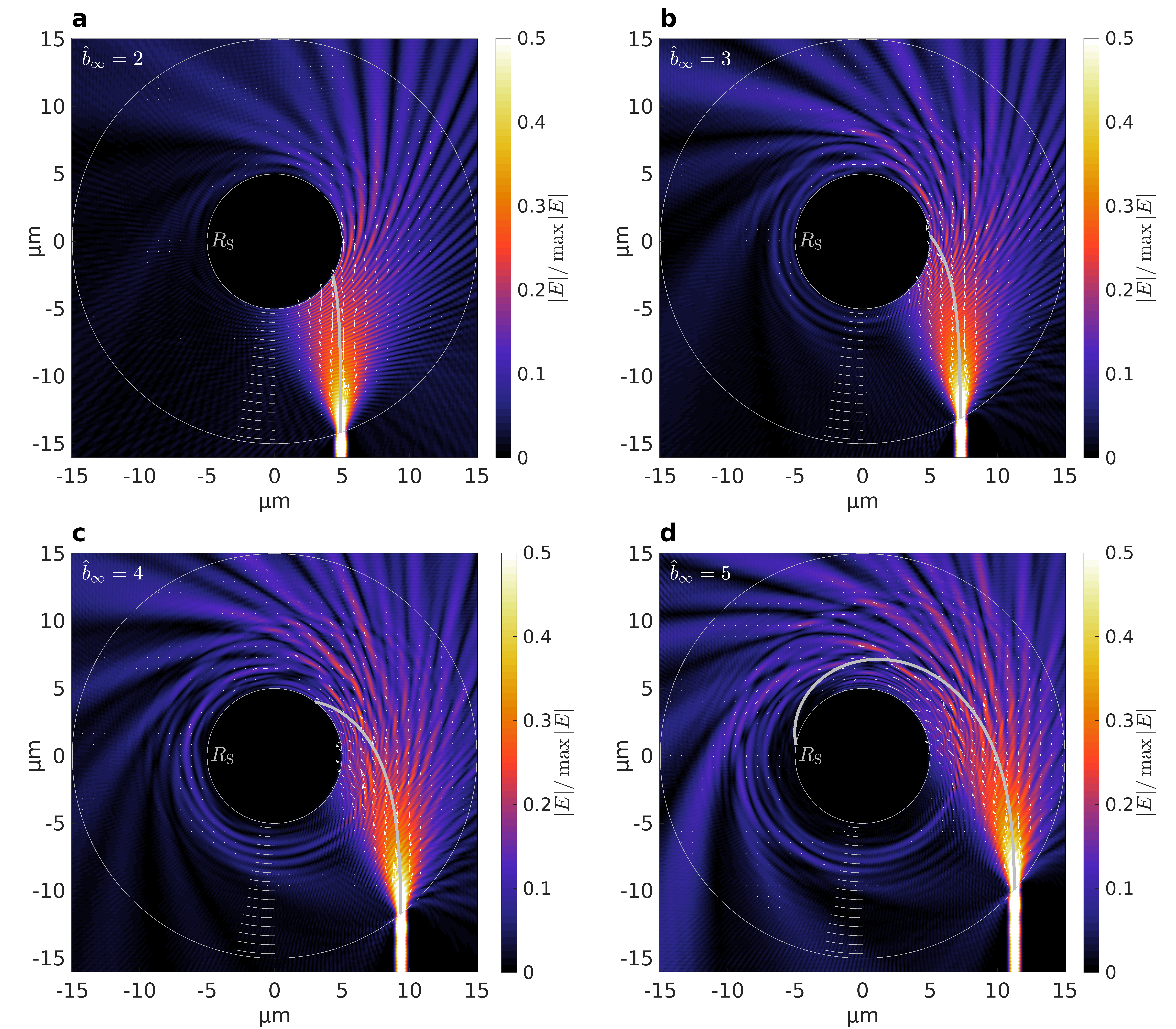}
            \caption{\final{\textbf{Numerical simulations of optical Schwarzschild black holes.}} Finite-difference frequency-domain simulations of light incident on an optical Schwarzschild black hole with impact parameters $\spacetimeImpactInfDD =$~\textbf{a}~2, \textbf{b}~3, \textbf{c}~4, and \textbf{d}~5. True geodesics are plotted as thick lines. Poynting vectors (white arrows) are scaled by $\propto1/\realR$. \final{Edge radii} and Schwarzschild radii \final{($\realSchwarzR$)} are solid circles. Each interior annulus's edge is marked. Color scales for \final{the normalized electric field amplitude} $\abs{\electricField} / \!\max\abs{\electricField}$ are the same for each subplot.}
            \label{fig:scanImpact}
        \end{figure}

        In this section, the trajectory of light around an optical black hole is modeled using a finite-difference frequency-domain (FDFD) solver \cite{shin2012,shin2015} of Maxwell's equations. \final{Simulation details are provided in the Methods section.} \Cref{fig:nSchwarzschild} shows the profiles $\refIndex(\realR)$ used when modeling light incident on an optical Schwarzschild black hole with four different impact parameters, $\spacetimeImpactInfDD = 2$, 3, 4, and 5; the resulting FDFD simulations are shown in \final{\cref{fig:scanImpact}, with the wavelength of light $\wavelength = \SI{0.5}{\um}$ and optical Schwarzschild radius $\realSchwarzR = 2 M = \SI{5}{\um}$.}

        Consider the simulation shown in \cref{fig:bInf10}, for which the dedimensionalized impact parameter at infinity is $\spacetimeImpactInfDD = 2$. The peak of the electric field normalized to its maximum, $\abs{\electricField} / \!\max\abs{\electricField}$, follows the path of the geodesic quite closely. Here, $\abs{\electricField} = \sqrt{\electricField \electricField^*}$, with $\electricField^*$ the complex conjugate of $\electricField$. Time-averaged Poynting vectors are calculated as $\re\left(\frac{1}{2}\V{\electricField}\times \V{\magnetizingField}^*\right)$ and scaled by $\propto1/\realR$ in the figures. Those with largest magnitude point mostly along the geodesic, and much of the energy flux is directed into the optical black hole. The same spatial trend is seen in \cref{fig:bInf15}, for which $\spacetimeImpactInfDD = 3$. Note in \cref{fig:nSchwarzschild} how the profile of scalar index $\refIndex(\realR)$ increases in amplitude as the initial impact parameter increases, in order to further bend light toward the horizon.

        For $\spacetimeImpactInfDD = 4$ and 5, seen in \cref{fig:bInf20,fig:bInf25}, respectively, the brightest regions of $\abs{\electricField} / \!\max\abs{\electricField}$ (and longest Poynting vectors) predominantly follow the geodesics. This is actually seen more clearly in the energy contained in the electric field ($\propto \abs{\electricField}^2$); however, only the electric field amplitude is shown here for better visualization of both small and large amplitude features. Agreement between the simulated light path and actual geodesic is expected to improve as the wavelength and beam width decrease relative to the size of the optical black hole, as described in \final{the following section}.

        Another interesting effect is observed in \cref{fig:bInf25}: the FDFD simulation does not show light following the geodesic all the way to the horizon. Instead, light begins to orbit at the photon sphere, $\realR = 3\bhMass = \SI{7.5}{\um}$. This results because the impact parameter is nearly equal to that at which light becomes trapped, $\spacetimeImpactInfDD = 3\sqrt{3} \approx 5.2$. Only traces of the photon ``ring'' are resolved in \cref{fig:bInf25}. Higher fidelity simulations, with the optical black hole comprising many more annuli, would likely be required to properly simulate and study this phenomenon. This is left to future work.

\begin{figure}[ht!]
            \centering
            \begin{subfigure}{\textwidth}
                \phantomcaption{}
                \label{fig:aHalfq0}
                \phantomcaption{}
                \label{fig:a0qHalf}
                \phantomcaption{}
                \label{fig:aPosFifth}
                \phantomcaption{}
                \label{fig:aNegFifth}
            \end{subfigure}
            \includegraphics[width=\textwidth]{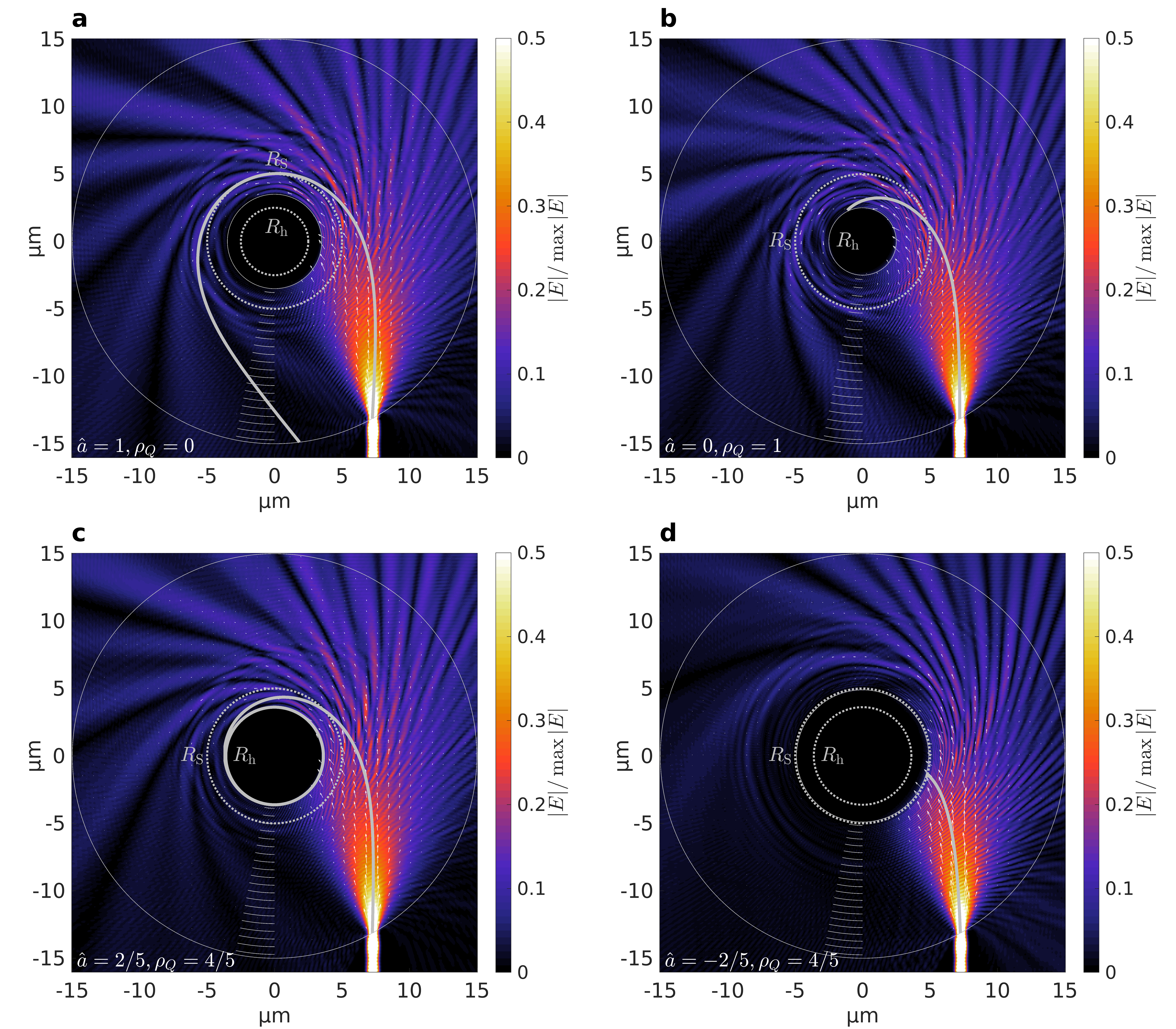}
            \caption{\final{\textbf{Numerical simulations of optical \KN{} black holes.}} Finite-difference frequency-domain simulations of light incident $(\spacetimeImpactInfDD = 3)$ on four optical \KN{} black holes, which are \textbf{a}~maximally co-rotating ($\knADD = 1, \knChargeRadiusDD = 0$); \textbf{b}~maximally charged ($\knADD = 0, \knChargeRadiusDD = 1$); \textbf{c}~charged and co-rotating ($\knADD = 2 / 5, \knChargeRadiusDD = 4 / 5$); and \textbf{d}~charged and counter-rotating ($\knADD = -2 / 5, \knChargeRadiusDD = 4 / 5$). True geodesics are plotted as thick lines. Poynting vectors (white arrows) are scaled by $\propto1/\realR$. Maximum and minimum radii are solid circles; radii of interest, such as the horizon \final{radius ($\realHorizonR$)} or Schwarzschild radius \final{($\realSchwarzR$)}, are also plotted and labeled. Each annulus's edge is marked. Color scales for \final{the normalized electric field amplitude} $\abs{\electricField} / \!\max\abs{\electricField}$ are the same for each subplot.}
            \label{fig:scanKerrNewman}
        \end{figure}

        Several optical \KN{} black holes are also simulated, with profiles $\refIndex(\realR)$ in \cref{fig:nKerrNewman} corresponding to the FDFD solutions in \cref{fig:scanKerrNewman}. For each case, the impact parameter is $\spacetimeImpactInfDD = 3$, and the outer edge of the optical black hole is again at radius $\realRDD_0 = 6$. These can be compared to the optical Schwarzschild black hole of \cref{fig:bInf15}. The innermost modeled radius varies for each simulation, depending on whether $\refIndex(\realRDD)$ diverges outside of the horizon. Each simulation is described in detail below.

        An extremal Kerr black hole ($\knADD = 1, \knChargeRadiusDD = 0$), with beam trajectory co-rotating with the black hole spin, is shown in \cref{fig:aHalfq0}. Here, $\refIndex(\realRDD)$ diverges at $\realRDD_* = 4 / 3$; however, the true geodesic escapes the black hole with $\D[\spacetimeR]{\affine} = 0$ at $\realRDD=2$. Therefore, though the horizon is at $\realHorizonRDD = 1$, the system is modeled with innermost radius $\realRDD = 1.4$. Comparing to the Schwarzschild case in \cref{fig:bInf15}, we see that light is ``dragged'' further around the co-rotating black hole, as expected. Additionally, more light ``escapes,'' although not all is directed along the geodesic. Inevitably, some energy flux is directed into the optical black hole, as indicated by the Poynting vectors; this is partially due to the finite width of the beam, and partially to the discrete annular approximation of the true gradient-index profile.

        In contrast to \cref{fig:aHalfq0}, an extremal \RN{} black hole ($\knADD = 0, \knChargeRadiusDD = 1$) is simulated, with FDFD results depicted in \cref{fig:a0qHalf}. Here, the optical black hole could be modeled completely to the horizon at $\realHorizonR = \SI{2.5}{\um}$. In general, the peak of $\abs{\electricField} / \!\max\abs{\electricField}$ follows the geodesic to the horizon. Little difference is seen when comparing to the Schwarzschild case of \cref{fig:bInf15}, except that light now propagates within $\realSchwarzR = \SI{5}{\um}$.

        Two non-extremal \KN{} black holes, with the same charge ($\knChargeRadiusDD = 4 / 5$) but opposite spins ($\knADD = \pm2 / 5$), are also simulated and shown in \cref{fig:aPosFifth,fig:aNegFifth}. The co-rotating black hole is modeled to the horizon at $\realHorizonRDD = 1 + \sqrt{1/5} \approx 1.45$. Compared to the extremal Kerr black hole in \cref{fig:aHalfq0}, light is not dragged as far around the black hole.

        For the counter-rotating \KN{} black hole, the profile $\refIndex(\realRDD)$ diverges at $\realRDD_* \approx 1.88$; this is the radius at which the geodesic begins co-rotating with the black hole spin, i.e., where $\D[\spacetimePhi]{\affine} = 0$. Thus, the system is modeled only to $\realRDD = 1.96$, where $\refIndex(\realRDD = 1.96) \approx 6$. Comparing the co- and counter-rotating black holes, we see that light travels further in the $\realPhi$-direction for the former system, as expected.

        As described in this section, a variety of optical Schwarzschild and \KN{} black holes can be constructed feasibly with low indices of refraction. If such systems are built at a small scale, FDFD simulations show that the trajectories of light behave as expected, mostly following the true geodesics despite the discrete approximation to the proper gradient-index profile. The benefits of building larger systems are discussed in the next section.

    \subsection*{Ray tracing calculations}\label{sec:rayTracing}
        In principle, the optical black holes of the previous section could be scaled in size from $\si{\um}$ to $\si{\cm}$ or larger. This would simplify not only the construction of the optical black hole but also the calculation of light propagation, since the wavelength and width of the light source would be much smaller than the system size and related gradient length scales. The minimum gradient scale length of the scalar-index profiles in \cref{fig:nSchwarzschild,fig:nKerrNewman} is $\refIndex / \norm{\grad\refIndex} \sim \SI{0.6}{\um}$, so $\wavelength < \refIndex / \norm{\grad\refIndex}$ is valid for the above FDFD simulations. If visible light, $\wavelength \approx 0.3-\SI{0.7}{\um}$, is used, scaling the system size by even a factor of $10^3$, i.e., from $\si{\um}$ to $\si{mm}$, or greater would be appropriate for the validity of the ray tracing approximation made in this section.

        It is of interest to calculate the deviation of a ray trajectory around the optical black hole from the true geodesic. These deviations could occur for a number of reasons: for instance, the discretization of $\refIndex(\realR)$ due to the finite number of annuli; manufacturing error, leading to an offset $\Delta\refIndex$ of the desired scalar index; or experimental error, resulting in a deviation $\Delta\realImpact_0$ from the desired initial impact parameter $\realImpact_0$. We explore the impacts of these below for light incident on an optical Schwarzschild black hole with outer radius $\realRDD_0 = 6$.

        First, we investigate the number of annuli (with uniform thicknesses) needed to sufficiently approximate the scalar-index profile for a range of initial impact parameters. We define our performance metric as the deviation of the ray trajectory from the geodesic, quantified by the difference in azimuthal angle $\Delta\realPhi = \realPhi_\text{ray} - \realPhi_\text{geo}$. \final{Note that this performance metric is design-specific and does not account for scattering, whereas the semi-classical calculations of \cite{narimanov2009,kildishev2010} do. However, as there is no analytic solution to the wave equation for the system under consideration, the pursuit of a more appropriate metric is left to future work.} Here, we are concerned with the deviation at the horizon. This value is shown in \cref{fig:nAnnuli} for $\spacetimeImpactInfDD \in [0,5]$ and the number of annuli ranging 1 to 50. We see that only 25 annuli are needed to reproduce trajectories with $\spacetimeImpactInfDD \leq 3$ to within $\Delta\realPhi = \SI{3}{\degree}$. As expected, $\Delta\realPhi$ increases rapidly for large $\spacetimeImpactInfDD$ and fixed annulus number. However, even the trajectory with $\spacetimeImpactInfDD = 5$ can achieve $\Delta\realPhi \leq \SI{3}{\degree}$ with 1000 annuli.

        \renewcommand{\tempWidth}{0.5\textwidth}
        \begin{figure}[ht!]
            \centering
            \includegraphics[width=\tempWidth]{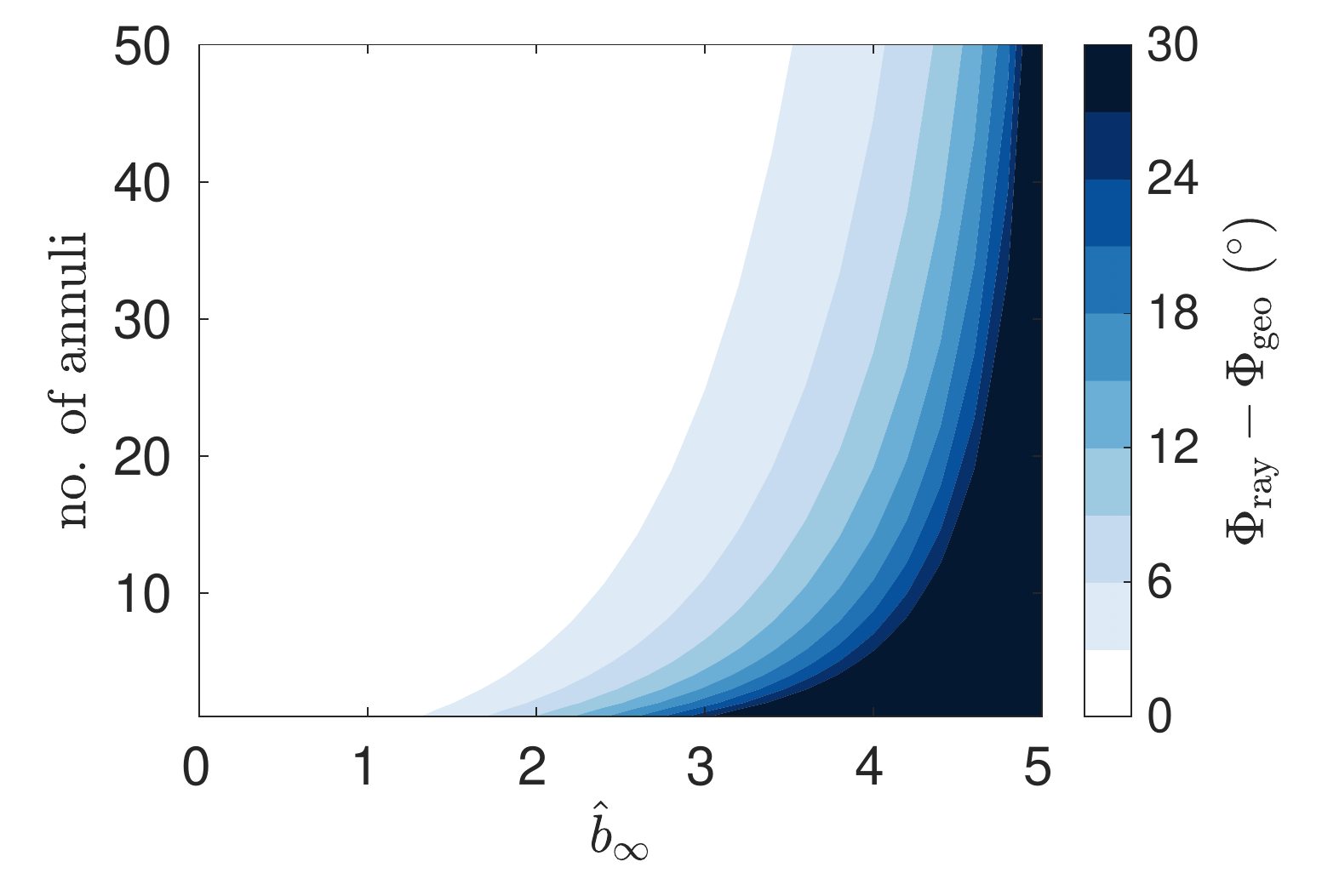}
            \caption{\final{\textbf{Impact of annulus number on ray trajectories.}} The angular deviation of the ray trajectory \final{($\realPhi_\text{ray}$)} from the geodesic \final{($\realPhi_\text{geo}$)} at the horizon for an optical Schwarzschild black hole with outer radius $\realRDD_0 = 6$, as a function of the initial impact parameter $\spacetimeImpactInfDD$ and number of annuli used in the construction.}
            \label{fig:nAnnuli}
        \end{figure}

        Next, we consider the scenario in which the scalar-index profile is imperfect, offset by a constant $\Delta\refIndex$ due to some manufacturing error. We choose a specific trajectory with impact parameter $\spacetimeImpactInfDD = 3$ to connect with the FDFD simulations of the previous section. The range spans $\Delta\refIndex \in [0, 0.5]$ in \cref{fig:rayTracing}; this is a significant percent change compared to profile~b in \cref{fig:nSchwarzschild}. \final{In \cref{fig:rayTracingA},}
        we see that the ray trajectory skews radially outward as $\Delta\refIndex$ increases. We are again interested in the deviation of the ray trajectory from the geodesic, $\Delta\realPhi = \realPhi_\text{ray} - \realPhi_\text{geo}$, \final{shown in \cref{fig:rayTracingB}} as a function of radius $\realRDD = \realR/\bhMass$. Most trajectories follow the geodesic closely, within $\Delta\realPhi \le \SI{2}{\degree}$, for $\realRDD > 3$; however, within $\realRDD < 3$, $\Delta\realPhi$ grows rapidly. The small grey region, near $\realRDD \approx 2$ and $\Delta\refIndex \approx 0.5$, indicates that the ray trajectory escapes the black hole, so that $\Delta\realPhi$ diverges. In this case, if errors of $\Delta\realPhi \le \SI{5}{\degree}$ were allowable, then $\refIndex(\realR)$ must be constrained with $\Delta\refIndex \le 0.1$.

        \renewcommand{\tempWidth}{0.9\textwidth}
        \begin{figure}[h!]
            \centering
            \begin{subfigure}{\tempWidth}
                \phantomcaption{}
                \label{fig:rayTracingA}
                \phantomcaption{}
                \label{fig:rayTracingB}
                \phantomcaption{}
                \label{fig:rayTracingC}
                \phantomcaption{}
                \label{fig:rayTracingD}
            \end{subfigure}
            \includegraphics[width=\tempWidth]{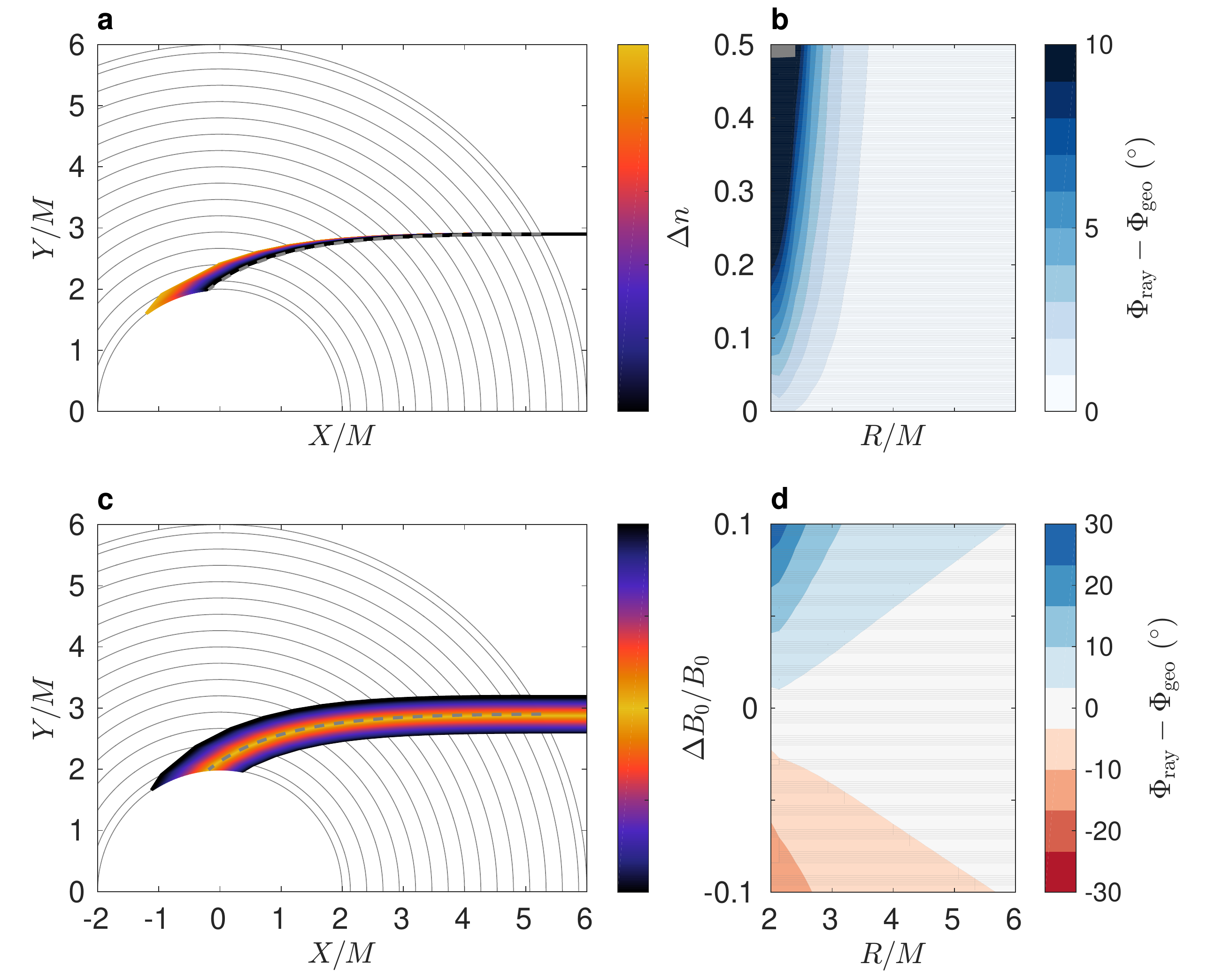}
            \caption{\final{\textbf{Effects of construction and experimenter errors on ray trajectories.} \textbf{a}, \textbf{b}: A uniform offset $\Delta\refIndex$ from the true scalar refractive index profile $\refIndex(\realR)$. \textbf{c}, \textbf{d}: A deviation $\Delta\realImpact_0$ from the desired impact parameter $\realImpact_0$. \textbf{a}, \textbf{c}: Ray trajectories in real space, computed from \cref{eq:rayTracing}, compared to the true geodesic (grey dashed). \textbf{b}, \textbf{d}: Angular deviation of the ray trajectory ($\realPhi_\text{ray}$) from the geodesic ($\realPhi_\text{geo}$) versus radius $\realR$ normalized to the black hole mass $\bhMass$. Both scans use the optical Schwarzschild black hole with $\spacetimeImpactInfDD = 3$. The scale of the color bar of subplot \textbf{a} is the same as the scale of the vertical axis of subplot \textbf{b}; the same is true for subplots \textbf{c} and \textbf{d}.}}
            \label{fig:rayTracing}
        \end{figure}

        In addition, a scan in initial impact parameter is performed to assess how experimental error would affect the ray trajectory. The ratio $\Delta\realImpact_0 / \realImpact_0$ is varied within $\pm 10\%$, with results shown in \cref{fig:rayTracing}. The ray trajectories \final{(\cref{fig:rayTracingC})} vary as expected: as $\abs{\Delta\realImpact_0}$ increases, the ray path moves farther from the true geodesic, but keeps the same general shape. Again, the deviation in azimuthal angle is shown \final{in \cref{fig:rayTracingD}.} For large $\abs{\Delta\realImpact_0} / \realImpact_0$, $\Delta\realPhi$ increases rapidly as the trajectory approaches the horizon. The deviation can be as large as $\Delta\realPhi = \SI{30}{\degree}$ at $\realRDD = 2$ when $\Delta\realImpact_0 / \realImpact_0 \approx 10\%$. Interestingly, the contours of $\Delta\realPhi$ versus $\realRDD$ and $\Delta\realImpact_0 / \realImpact_0$ are not symmetric about $\Delta\realImpact_0 / \realImpact_0 = 0$ in \final{\cref{fig:rayTracingD}}. This results from the discretization of $\refIndex(\realR)$. Therefore, if the number of annuli cannot be increased, it could actually be beneficial to purposefully shift the impact parameter ($\Delta\realImpact_0 / \realImpact_0 < 0$, in this case) to better match the light trajectory with the true geodesic.

\section*{Discussion}\label{sec:conclusion}
    The application of analogue spacetimes to \final{the} study of general relativity has seen a resurgence in theory, simulation, and experiment in the past two decades. Many recent works have focused on optical analogues to static, uncharged (Schwarzschild) black holes in an isotropic coordinate system. In this paper, we have calculated the dielectric permittivity and permeability tensors $\permittivityTensor, \permeabilityTensor$ that reproduce the equatorial null geodesics and polarizations of light moving in the metric of spinning, charged (\KN{}) black holes. Furthermore, we have conceived, for the first time, a gradient-index material that exactly reproduces families of equatorial \KN{} null geodesics in almost all cases. Importantly, the radial profile of the scalar refractive index $\refIndex(\realR)$ is finite along the entire trajectory (even to the horizon, if applicable), except at the point of rotation reversal for initially counter-rotating null geodesics. Values of $\refIndex \lesssim 6$ can be achieved for many trajectories of interest, meaning that such gradient-index optical analogues could be constructed with conventional materials and metamaterials.

    Simulations of a variety of optical black holes were performed, each with $\refIndex(\realR)$ approximated by concentric circular annuli of constant scalar index. First, a finite-difference frequency-domain (FDFD) solver of Maxwell's equations was used to simulate the path of light incident on a Schwarzschild black hole with varying impact parameter $\spacetimeImpactInfDD = \spacetimeImpactInf / \bhMass$. Good agreement was observed between the light trajectory (indicated by maximum values of the electric field and Poynting vectors) and geodesic for low impact parameters $\spacetimeImpactInfDD = \text{2--3}$, but the discrepancy grew for $\spacetimeImpactInfDD = \text{4--5}$. Interestingly, for $\spacetimeImpactInfDD = 5$, some features of light orbiting at the photon sphere were observed. Utilizing the same FDFD framework, several optical \KN{} black holes were simulated: extremal Kerr, extremal \RN{}, and non-extremal \KN{} with initially co- and counter-rotating trajectories. Each of these optical systems was simulated within the Schwarzschild radius, some even to the horizon. While there exist some discrepancies between the simulated light trajectories and true geodesics, the qualitative feature of light ``dragged'' in the direction of the black hole's spin was observed. The three co-rotating cases require $\refIndex \lesssim 3$, meaning that constructions of these optical \KN{} black holes are feasible; the counter-rotating case requires $\refIndex \lesssim 6$, which might be realized with more exotic materials like metamaterials.

    Finally, we have investigated the number of annuli used in construction as well as the effects of fabrication and experimental errors on these optical black holes. \final{The results demonstrate that with a modest number of annuli, the approximate gradient-index systems adequately reproduce null geodesics and are robust to small variations in refractive index and impact parameter. As these systems are far easier to manufacture than true gradient-index or bianisotropic media, they are thus practical tabletop analogues for equatorial \KN{} black holes.}

\section*{Methods}\label{sec:methods}
    \subsection*{Numerical simulations}\label{sec:simulations}
        \final{
        The trajectories of light around an optical black holes are modeled using a finite-difference frequency-domain (FDFD) solver of Maxwell's equations \cite{shin2012,shin2015}. The wavelength of light is chosen to be $\wavelength = \SI{0.5}{\um}$. The 2D simulation domain is modeled as a vacuum, with scalar properties $\permittivityTensor = \permeabilityTensor = \refIndex = 1$ and size $60 \wavelength \times 60 \wavelength$; a perfectly matching layer of width $\wavelength / 5$ is applied at its boundary. A Gaussian beam of light is approximated as an array of line sources, each of width $\wavelength/25 = \SI{20}{\nm}$ and electric field amplitude calculated from a Gaussian envelope of the form $\exp(-(\realX - \realImpact_0)^2 / 2 \GaussWidth^2)$. Here, $\realImpact_0$ is the dimensionful real space impact parameter at $\realRDD_0$, and $\GaussWidth = \wavelength / 2$ so that the beam satisfies the paraxial approximation \cite{nemoto1990}. The total width of the beam is truncated at $2 \wavelength$ by imposing two absorbing ($\permittivityTensor = 1 - i \pi$) boundaries as vertically aligned ``waveguides'' of the light from the edge of the domain to the edge of the optical black hole. These restrict the beam to travel along a straight path in free space, as a directional light source would in the laboratory. Note that the factor of $-\pi$ is arbitrarily chosen for the imaginary (damping) component.
        }

        \final{
        Each simulated optical black hole is centered in the domain, with the Schwarzschild radius always $\realSchwarzR = 10 \wavelength = \SI{5}{\um}$ ($\bhMass = \SI{2.5}{\um}$) and edge at $\realR_0 = 30 \wavelength = \SI{15}{\um}$. The Gaussian light source propagates in the vertical direction toward the black hole. For all simulations, the region within the minimum radius (oftentimes the horizon radius $\realHorizonR$) is modeled as a disc with dielectric permittivity $\permittivityTensor = \permittivityTensor_{\mathrm{in}} - i\pi$. Here, $\permittivityTensor_{\mathrm{in}}$ is the scalar permittivity ($\permittivityTensor = \refIndex^2$) of the innermost annulus, and a factor of $-\pi$ is used for the imaginary (damping) component, as with the aforementioned ``waveguides.''
        }

    \subsection*{Ray tracing algorithm}\label{sec:raytracing}
        \final{
        Consider an optical system consisting of $N$ concentric annuli. Let the radii bounding each annulus $i$ be $\realR_i < \realR_{i - 1}$, so that the annuli are numbered $1, 2, \dots, N$ from the outside in, and the outer edge of the system is at $\realR_0$. The scalar index of each annulus is $\refIndex(\realR_i < \realR \le \realR_{i - 1}) = \refIndex_i$, which monotonically increases from annulus $1\to N$, so $\refIndex_i < \refIndex_{i+1}$. Let the scalar index for $\realR > \realR_0$ be $\refIndex_0$. For a light ray incident on annulus ($i + 1$) (propagating in the region $\realR_i \le \realR \le \realR_{i - 1}$), let the impact parameter be $\realImpact_i = \realR_i \sin\realPhi_i$, where $\realPhi_i$ is the azimuthal angle at which the ray intersects the annulus at $\realR_i$. Then, the azimuthal angle at which the light ray intersects the next annulus ($i + 2$) at $\realR_{i + 1}$ is given by
            \begin{equation}
                \label{eq:rayTracing}
                \realPhi_{i + 1} - \realPhi_i = \arcsin\parens*{\frac{\realImpact_{i + 1}}{\realR_{i + 1}}} - \arcsin\parens*{\frac{\realImpact_{i + 1}}{\realR_i}}\,,
            \end{equation}
        provided that $\realImpact_{i + 1} \le \realR_{i + 1}$. Note that the impact parameter always satisfies $\refIndex_i\realImpact_i = \text{constant}$. Thus, given an optical system with a well-defined profile $\refIndex(\realR)$ and an initial impact parameter $\realImpact_0$, the trajectory of a light ray can be iteratively computed via \cref{eq:rayTracing} until the ray reaches its minimum radius. Note that only in-going trajectories are considered here, so light escaping the optical black hole is not modeled. Furthermore, it is assumed that all light is transmitted at each boundary; absorption and reflection are left for future work.
        }

    \subsection*{Derivation of \KN{} analogue material properties}\label{sec:derivation}
        \final{
        Here, we derive \cref{eq:kerrTens}, beginning with \cref{eq:analogueEpsMuAl,eq:ddBLMetric}. Restricting our attention to equatorial geodesics, we have $\spacetimeTheta = \pi / 2$ and $\diff\spacetimeTheta = 0$ (as the motion will always remain equatorial). With this, the metric simplifies to
            \begin{equation}
                \label{eq:ddBLMetricEquatorial}
                \diff\spacetimeArcLengthDD^2 = \spacetimeRDD^2 \parens*{\frac{\diff\spacetimeRDD^2}{\knDeltaDD} + \diff\spacetimeTheta^2} - \frac{\knDeltaDD}{\spacetimeRDD^2} \parens*{\diff\spacetimeTimeDD - \knADD \diff\spacetimePhi}^2  + \frac{1}{\spacetimeRDD^2} \bracks*{\parens*{\spacetimeRDD^2 + \knADD^2} \diff\spacetimePhi - \knADD \diff\spacetimeTimeDD}^2\,.
            \end{equation}
        Expanding this, we find
            \begin{equation}
                \metricTensor_{\mu \nu} =
                    \mat[p]{
                        -\frac{\knDeltaDD - \knADD^2}{\spacetimeRDD^2} & 0 & 0 & \frac{\knADD \parens*{\knDeltaDD - \spacetimeRDD^2 - \knADD^2}}{\spacetimeRDD^2} \\
                        0 & \frac{\spacetimeRDD^2}{\knDeltaDD} & 0 & 0 \\
                        0 & 0 & \spacetimeRDD^2 & 0 \\
                        \frac{\knADD \parens*{\knDeltaDD - \spacetimeRDD^2 - \knADD^2}}{\spacetimeRDD^2} & 0 & 0 & \frac{\parens*{\spacetimeRDD^2 + \knADD^2}^2 - \knADD^2 \knDeltaDD}{\spacetimeRDD^2}
                    }\,,
            \end{equation}
        where $\mu, \nu$ run over $\spacetimeTimeDD, \spacetimeRDD, \spacetimeTheta, \spacetimePhi$. This metric has inverse
            \begin{equation}
                \metricTensor^{\mu \nu} =
                    \mat[p]{
                        \frac{\knADD^2 \knDeltaDD - \parens*{\spacetimeRDD^2 + \knADD^2}^2}{\spacetimeRDD^2 \knDeltaDD} & 0 & 0 & \frac{\knADD \parens*{\knDeltaDD - \spacetimeRDD^2 - \knADD^2}}{\spacetimeRDD^2 \knDeltaDD} \\
                        0 & \frac{\knDeltaDD}{\spacetimeRDD^2} & 0 & 0 \\
                        0 & 0 & \frac{1}{\spacetimeRDD^2} & 0 \\
                        \frac{\knADD \parens*{\knDeltaDD - \spacetimeRDD^2 - \knADD^2}}{\spacetimeRDD^2 \knDeltaDD} & 0 & 0 & \frac{\knDeltaDD - \knADD^2}{\spacetimeRDD^2 \knDeltaDD}
                    }
            \end{equation}
        and determinant $\det\metricTensor = -\spacetimeRDD^4$. We map the curved spacetime coordinates $(\spacetimeTimeDD, \spacetimeRDD, \spacetimeTheta, \spacetimePhi)$ onto the flat spacetime spherical coordinates $(\realTimeDD, \realRDD, \realTheta, \realPhi)$, so the flat space coordinate metric is in this case
            \begin{equation}
                \coordMetric_{i j} =
                    \mat[p]{
                        1 & 0 & 0 \\
                        0 & \realRDD^2 & 0 \\
                        0 & 0 & \realRDD^2 \sin^2\realTheta
                    }
            \end{equation}
        with determinant $\det\coordMetric = \realRDD^4 \sin^2\realTheta$. Because we have restricted our attention to $\spacetimeTheta = \pi / 2$, we similarly have $\realTheta = \pi / 2$, and so this simply becomes $\det\coordMetric = \realRDD^4$. After this coordinate matching, we have
            \begin{equation}
                \begin{aligned}
                    \metricTensor^{i j} &= \mat[p]{
                        \frac{\knDeltaDD}{\realRDD^2} & 0 & 0 \\
                        0 & \frac{1}{\realRDD^2} & 0 \\
                        0 & 0 & \frac{\knDeltaDD - \knADD^2}{\realRDD^2 \knDeltaDD}
                    }\,, \\
                    \metricTensor_{0 0} &= -\frac{\knDeltaDD - \knADD^2}{\realRDD^2}\,, \\
                    \metricTensor_{0 i} &= \mat[p]{0 & 0 & \frac{\knADD \parens*{\knDeltaDD - \realRDD^2 - \knADD^2}}{\realRDD^2}}\,, \\
                    \det\metricTensor   &= -\realRDD^4\,,
                \end{aligned}
            \end{equation}
        where $\knDeltaDD$ is now interpreted as a function of $\realRDD$, as opposed to $\spacetimeRDD$. Plugging these values into \cref{eq:analogueEpsMuAl}, we arrive at
            \begin{equation}
                \permittivityTensor^{i j} = \permeabilityTensor^{i j} = \mat[p]{\frac{\knDeltaDD}{\knDeltaDD - \knADD^2} & 0 & 0 \\ 0 & \frac{1}{\knDeltaDD - \knADD^2} & 0 \\ 0 & 0 & \frac{1}{\knDeltaDD}}\,, \quad \magnetoelectric_i = \mat[p]{0 \\ 0 \\ \knADD \parens*{\frac{1}{\knDeltaDD - \knADD^2} - \frac{1}{\realRDD^2}}}\,.
            \end{equation}
        }

\section*{Data availability}\label{sec:dataAvailability}
    Data is available from the corresponding author upon request.

\section*{Code availability}\label{sec:codeAvailability}
    Code for the simulations shown here is available from the corresponding author upon request. The finite-difference frequency-domain solver used in this work is available at\newline \href{https://github.com/wsshin/maxwellfdfd}{\texttt{https://github.com/wsshin/maxwellfdfd}}.

\section*{Acknowledgments}\label{sec:acknowledgments}
    This work grew from a submission to the Harvard Black Hole Initiative essay competition. The authors thank K. R. Moore for inspiration; R. Bekenstein, S. G. Johnson, N. Rivera, and S. P. Robinson for fruitful discussions; A. Patterson and the MIT Department of Physics. A.P.T. gratefully acknowledges the Tushar Shah and Sara Zion Fellowship for funding and was partially supported by DOE grant DE-SC00012567. The authors are also grateful to the MIT Open Access Article Publication Subvention Fund, the MIT Plasma Science and Fusion Center, \final{and the MIT Center for Theoretical Physics for their support}.



\bibliographystyle{unsrt}
\bibliography{bib2}

\end{document}